\documentclass[runningheads]{llncs}
\usepackage{graphicx}
\usepackage{cite}
\usepackage{amsmath,amssymb,amsfonts}
\usepackage{algorithmic}
\usepackage{graphicx}
\usepackage{textcomp}
\usepackage{xcolor}

\usepackage[utf8]{inputenc}
\usepackage{colortbl,array} % für farbige cells
\usepackage{multirow,bigdelim}
\usepackage{amsmath}
\usepackage{placeins}
\usepackage{stfloats}
\usepackage{amsfonts}
\usepackage{booktabs}
\usepackage{multirow}
\usepackage[utf8]{inputenc}
\usepackage{graphicx}
\usepackage{tabularx}
\newcolumntype{L}{>{\raggedright\arraybackslash}X}

\begin{document}
	\title{Modality Completion via Gaussian Process Prior Variational Autoencoders for Multi-Modal Glioma Segmentation}
	\titlerunning{Modality Completion via GP for Multi-Modal Glioma Segmentation}
	% index{Last Name, First Name}
	\author{Mohammad Hamghalam\inst{1,2} \and % index{Hamghalam, Mohammad}
	Alejandro F. Frangi\inst{4,5,6} \and % index{Frangi, Alejandro F.}
	Baiying Lei\inst{7} \and   % index{Lei, Baiying}
	Amber L. Simpson\inst{1,3}} % index{Simpson, Amber L.}

%/Department of Biomedical and Molecular Sciences,
	
	\authorrunning{M. Hamghalam et al.}
	\institute{School of Computing, Queen’s University, Kingston, ON, Canada
		\email{m.hamghalam@gmail.com} \and
	Department of Electrical, Biomedical, and Mechatronics Engineering, Qazvin Branch, Azad University, Qazvin, 34116846-13114, Iran \and
	Department of Biomedical and Molecular Sciences, Queen's University, Kingston, ON, Canada \and	
	CISTIB Centre for Computational Imaging \& Simulation Technologies in Biomedicine, School of Computing, University of Leeds, Leeds LS2 9LU, UK 
	\and
	LICAMM Leeds Institute of Cardiovascular and Metabolic Medicine, School of Medicine, Leeds LS2 9LU, UK
	 \and
	Medical Imaging Research Center (MIRC) – University Hospital Gasthuisberg, KU Leuven, Herestraat 49, 3000 Leuven, Belgium \and
	National-Regional Key Technology Engineering Laboratory for Medical Ultrasound, Guangdong Key Laboratory for Biomedical Measurements and Ultrasound Imaging, School of Biomedical Engineering, Health Science Center, Shenzhen University, Shenzhen, China.
}
	\maketitle              % typeset the header of the contribution
	\begin{abstract}
		In large studies involving multi protocol Magnetic Resonance Imaging (MRI), it can occur to miss one or more sub-modalities for a given patient owing to poor quality (e.g. imaging artifacts), failed acquisitions, or hallway interrupted imaging examinations. In some cases, certain protocols are unavailable due to limited scan time or to retrospectively harmonise the imaging protocols of two independent studies. Missing image modalities
		pose a challenge to segmentation frameworks as complementary information contributed by the missing scans is then lost.
		In this paper, we propose a novel model, Multi-modal Gaussian Process Prior Variational Autoencoder (MGP-VAE), to impute one or more missing sub-modalities for a patient scan. MGP-VAE can leverage the Gaussian Process (GP) prior on the Variational Autoencoder (VAE) to utilize
		the subjects/patients and sub-modalities correlations. Instead of designing one network for each possible subset of present sub-modalities or using frameworks to mix feature maps, missing data can be generated from a single model based on all the available samples. We show the applicability of MGP-VAE on brain tumor segmentation where either, two, or three of four sub-modalities may be missing.
		Our experiments against competitive segmentation baselines with missing sub-modality on BraTS'19 dataset indicate the effectiveness of the MGP-VAE model for segmentation tasks. 
				
	\end{abstract}

	\section{Introduction}
	Glioma tumor segmentation in MR scans plays a crucial role during the diagnosis, survival prediction, and brain tumor surgical planning. 
	Multiple MRI sub-modalities, FLAIR (F), T1, T1c, and T2, are regularly utilized to detect and evaluate the brain tumor subregions such as the whole tumor (WT), tumor core (TC), and the enhancing tumor (ET) region. These sequences provide comprehensive information regarding tumor brain tissues.
	In clinical settings, it is common for physicians to have one or more sub-modalities to be missing for a patient due to patient artifacts, acquisition problems, and other clinical reasons. 
	
	Segmentation with missing modalities techniques can be categorized 
	into three approaches:
	1) training a segmentation model for any subset of input sub-modalities;
	2) training a synthesis model to impute missing sequence from input sub-modalities \cite{li2014deep};
	3) instead of designing different models for every potential missing modality combination, designing a single model that operates based on the shared feature space through all input sub-modalities (such as taking the mean) \cite{havaei2016hemis,varsavsky2018pimms,dorent2019hetero}.
	The two first solutions associate with the training and handling of a different network for $2^{(\# of\ sub-modalities)}-1$ combinations. The third group extracts a shared feature space from the sub-modalities, which is independent of the number of sub-modalities, to provide a unique model that shares its extracted features.
	
	The current methods, which work based on the common representation to address missing modality, are Hetero-modal Image Segmentation (HeMIS) \cite{havaei2016hemis} and its relevant extension Permutation Invariant Multi-Modal Segmentation (PIMMS) technique \cite{varsavsky2018pimms}. They computed first and second moments of extracted feature maps across available sub-modalities to combine them separately. Although using these statistics are independent of the number of sub-modalities, they do not compel their convolutional model to learn a shared latent representation. For this aim, Dorent \textit{et al.} \cite{dorent2019hetero} introduced Hetero-Modal Variational Encoder-Decoder (HVED) based on the Variational Autoencoder (VAE) \cite{kingma2013auto} to provide the common latent variable $z$. Furthermore, conditional VAE (CVAE) \cite{sohn2015learning} includes extra data in both the decoder and the encoder to produce samples with speciﬁc properties. Based on this method, some models used the auxiliary information to synthesis missing sub-modalities \cite{sharma2019missing}.
	 However, the prior assumption that latent representations are identically and independently distributed (\textit{i.i.d.})  is one of the VAE model's limitations.	 
	   In contrast, the Gaussian Process Prior Variational Autoencoder (GPPVAE) \cite{casale2018gaussian} models correlation between input sub-modalities through a Gaussian Process (GP) prior on the latent representations.
	
	In this paper, we extend the GPPVAE for missing MRI sub-modalities imputation in a 3D framework for brain glioma tumor segmentation. The contribution of this work is three-fold. First, we extend the GPPVAE for 3D MRI modalities imputation for any scenario of missing sequences as input in one model. Second, we adapt a kernel function to capture multiple levels of correlation between sub-modalities and subjects in our Multi-modal Gaussian Process Prior Variational Autoencoder (MGP-VAE). Finally, we show that our model outperforms HVED and HeMIS in terms of DSC from multi-modal brain tumor MRI scans with any configuration of the available sub-modalities.
	\section{Multi-modal Gaussian Process Prior Variational Autoencoder}
\begin{figure}[!t]
		\centering
		\includegraphics[width=.90\columnwidth]{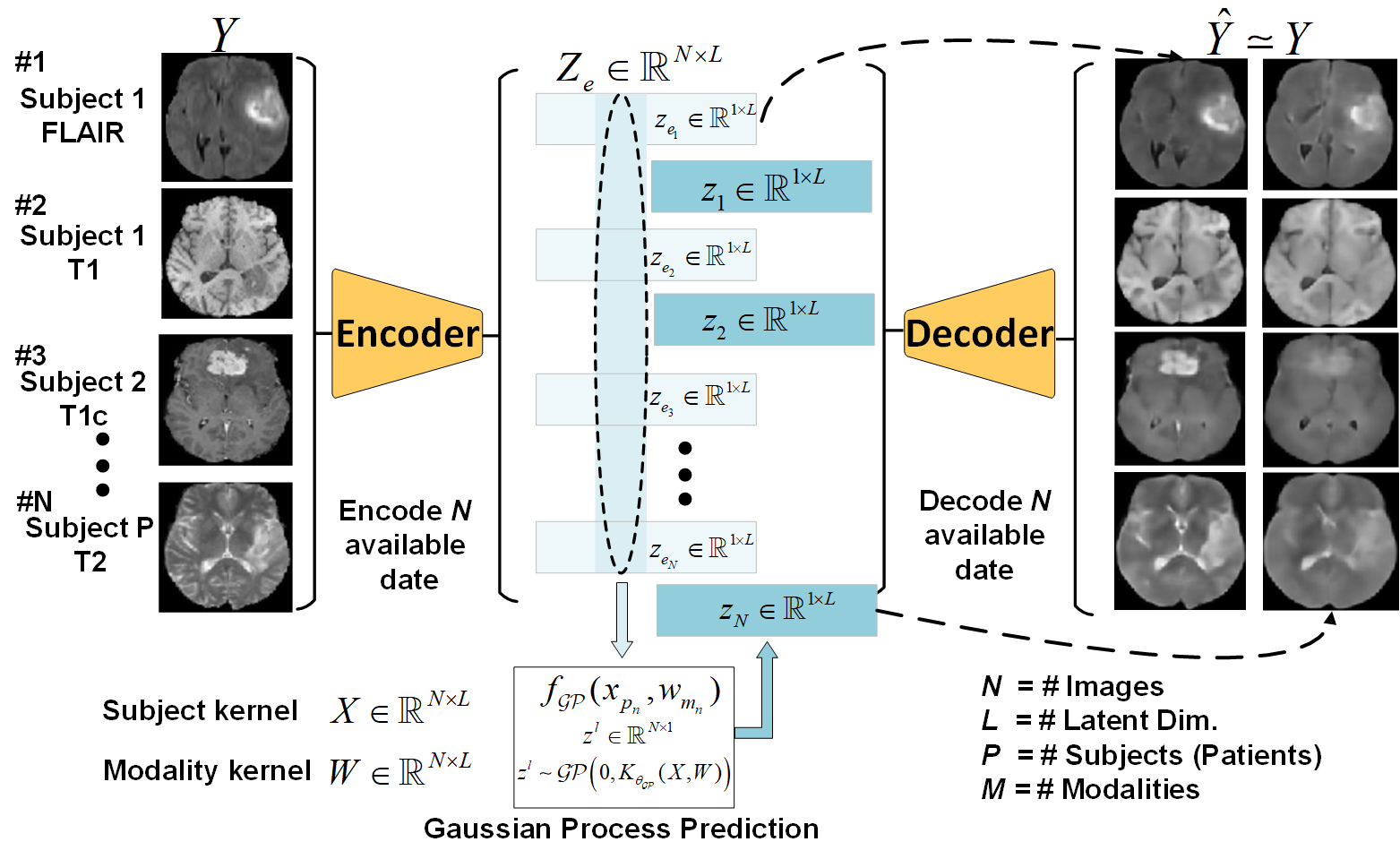} 
		\vspace{-.3cm}
		\caption{Overview of the proposed MGP-VAE. Each sub-modality volume is mapped to a 1024-dimensional ($L$) space and decoded to the initial space. Covariances among input volumes are formed through a GP Prior to each column of the latent representation matrix ${\textbf{Z}}_{e}$. The subject and modality correlations are modeled in the latent space due to its compact superiority.}
		\label{fig:overview}
	\end{figure}

	%\subsection{}
	%
	%
	%
	Assume we have a collection of 3D MRI scans to visualize brain and tumor tissue in different contrasts in MR pulse sequences, each scan coupled with auxiliary data: patient/subject IDs and sub-modality entities. Individually, we consider BraTS'19 datasets with MRI scans of the brain with glioma tumors in four sub-modalities per patient: F, T1, T1c, and T2, each of which contributes differing tissue contrast views and spatial resolutions. Each unique patient and sub-modality is assigned to a feature vector, which we define as a subject feature vector and a modality feature vector, respectively. Subject and modality features refer to elements of kernel covariance function calculated based on present data.
	Subject feature vectors might contain brain features such as brain tissue, tumor location, or dimension, while modality feature vectors may contain contrast information or tumor sub-region features of each modality. Fig. \ref{fig:overview} shows an overview of the MGP-VAE for synthesizing missing modalities. We assume that at least one sub-modality of each test subject is present during training.

\textbf{Formulation}.
	Let $P$ denote the number of subjects, $M$ the number of sub-modalities, $N=P\times M$ denote the number of all samples, and consider $\textbf{Y}=\{{{y}_{n}}\}_{n=1}^{N}\in {{\mathbb{R}}^{N\times K}}$ for all $N$ input samples and $K$ denotes $k$-dimensional representation for $N$ samples; let $\textbf{X}=\{{{x}_{p}}\}_{p=1}^{P}\in {{\mathbb{R}}^{P\times Q}}$ denote $Q$-dimensional patient feature vectors for the $P$ patients, and let $\mathbf{W}=\{{{w}_{m}}\}_{m=1}^{M}\in {{\mathbb{R}}^{M\times R}}$ denote $R$-dimensional modality feature vectors for the $M$ sub-modalities. Four sub-modalities provide complementary information about the brain tissue, $\textbf{W}=\{{{w}_{1}}=F,{{w}_{2}}=T1,{{w}_{3}}=T1c,{{w}_{4}}={{T}_{2}}\}$. Finally, let $\textbf{Z}=\{{{z}_{n}}\}_{n=1}^{N}\in {{\mathbb{R}}^{N\times L}}$ denote the $L$-dimensional latent representations which abstract input samples through GP, ${f}_{\mathcal{GP}}$. We examine the following process for the available input data:
	
	\begin{itemize}
	\item The latent representation of MRI scan ${{p}_{n}}$ in sub-modality ${{m}_{n}}$ is generated from the subject feature vector ${{x}_{{{p}_{n}}}}$ and modality feature vector ${{w}_{{{m}_{n}}}}$ as: ${{z}_{n}}={{f}_{\mathcal{GP}}}({{x}_{{{p}_{n}}}},{{w}_{{{m}_{n}}}})$,
	%
	%\begin{equation}\label{eq1}
	%\begin{aligned}
	%{{z}_{n}}={{f}_{\mathcal{GP}}}({{x}_{{{p}_{n}}}},{{w}_{{{m}_{n}}}})
	%\end{aligned}
	%\end{equation}
	%
	where ${f}_{\mathcal{GP}}$ is a GP prior to compute sample covariances as a function of subject and modality feature vectors in the latent space, $z$.

	\item Reconstructed output ${\hat{y}_{n}}$ is created from its latent representation ${{z}_{n}}$ as: ${{\hat{y}}_{n}}={{f}_{d}}({{z}_{n}})$,
	%
	%\begin{equation}\label{eq2}
%	\begin{aligned}
	%{{\hat{y}}_{n}}={{f}_{d}}({{z}_{n}})
	%\end{aligned}
	%\end{equation}
	%
	where ${f}_{d}$ is a convolutional neural network with decoder architecture to map latent representation, $z$, into the reconstruction space, $\hat{y}$.
	\end{itemize}
	The marginal likelihood of the MGP-VAE model is:
	\begin{equation}\label{eq3}
	\begin{aligned}
	p(\textbf{Y}|\textbf{X},\textbf{W},{{\theta }_{d}},{{\theta }_{\mathcal{G}\mathcal{P}}})=\int{p(\textbf{Y}|\textbf{Z},{{\theta }_{d}})} \ p(\textbf{Z}|\textbf{X},\textbf{W},{{\theta }_{\mathcal{G}\mathcal{P}}})\ d\textbf{Z}
	\end{aligned}
	\end{equation}
	where ${\theta}_{d}$ denotes the parameters of the decoder and ${\theta}_{\mathcal{GP}}$ indicates the  parameters of GP's kernels. Eq.\ref{eq3} cannot be optimized straightforwardly as the integral is not tractable. Thus we resort to variational inference, which requires introducing an approximate posterior distribution.
	\subsection{Proposed Kernel Functions for MGP-VAE}
	The GP defines a set of random variables on ${z}^{l}$, $l-$th column of $\textbf{Z}$, so any finite number of them have a multivariate Gaussian distribution. In case ${{z}^{l}}={{f}_{\mathcal{G}\mathcal{P}}}(\textbf{X},\textbf{W})$ is a GP, then given $L$ observations, the joint distribution of the random variables, ${{z}^{1}}={{f}_{\mathcal{G}\mathcal{P}}}(\textbf{X},\textbf{W}),{{z}^{2}}={{f}_{\mathcal{G}\mathcal{P}}}(\textbf{X},\textbf{W}),...,{{z}^{L}}={{f}_{\mathcal{G}\mathcal{P}}}(\textbf{X},\textbf{W})$, is Gaussian. For our $L-$dimensional latent representation, we have:
	\begin{equation}\label{eq4}
	\begin{aligned}
	p(\textbf{Z}|\textbf{X},\textbf{W},{{\theta }_{\mathcal{G}\mathcal{P}}})=\prod\limits_{l=1}^{L}{\mathcal{G}\mathcal{P}({{z}^{l}}|0,{{K}_{{{\theta }_{\mathcal{G}\mathcal{P}}}}}(\textbf{X},\textbf{W}))}
	\end{aligned}
	\end{equation}
	where ${{K}_{{{\theta }_{\mathcal{G}\mathcal{P}}}}}(\textbf{X},\textbf{W})$ is the covariance function with the kernel parameters, ${{\theta }_{\mathcal{G}\mathcal{P}}}$, which comprises a modality kernel and a patient kernel. The former models covariance among sub-modalities, while the latter models covariance between patients. 
	${{K}_{{{\theta }_{\mathcal{G}\mathcal{P}}}}}(\textbf{X},\textbf{W})$ can be factorized into \cite{bonilla2007kernel}:
	\begin{equation}\label{eq5}
	\begin{aligned}
	{{K}_{{{\theta}_{\mathcal{G}\mathcal{P}}}}}(\textbf{X},\textbf{W})=\underbrace{\mathcal{K}({{x}_{p}},{{{{x}'}}_{p}})}_{^{patient\ kernel}} \otimes \underbrace{\mathcal{K}({{w}_{m}},{{{{w}'}}_{m}})}_{^{modality\ kernel}}
	\end{aligned}
	\end{equation}
	where ${{x}_{p}}$ and ${{{x}'}_{p}}$ are feature vectors of two patients, ${{w}_{m}}$ and ${{{w}'}_{m}}$ are corresponding modality feature vectors. Also, $\otimes$ is the Kronecker product of these two matrices to make the dimensions match between the $P\times P$ and $M\times M$ matrix. These features are extracted from the latent space during training. We define $L = 1024$ as the latent space dimension. We set
	a full-rank covariance as a modality covariance ($\mathcal{K}({{w}_{m}},{{{w}'}_{m}})$) for our limited sub-modalities (F, T1, T1c, and T2) and a linear covariance (${\mathcal{K}({{x}_{p}},{{{{x}'}}_{p}})} = x_{p}^{T}.{{{x}'}_{p}}$) to measure similarity among the subjects with $Q$ = 64.
	
	\subsubsection{Loss Function and Optimization.}	
	As a standard VAE, we approximate the latent variables by a Gaussian distribution    whose mean and diagonal covariance are defined by two functions, ${\mu }_{e}({y}_{n})$ and $diag(\sigma _{e}^{2}({{y}_{n}}))$. Thus, we have:
	\begin{equation}\label{eq6}
	\begin{aligned}
	q({{\textbf{Z}}_{e}}|\textbf{Y})=\prod\limits_{n=1}^{N}{\mathcal{N}\left( {{z}_{{e}_{n}}}|{{\mu }_{e}}({{y}_{n}}),diag(\sigma _{e}^{2}({{y}_{n}})) \right)},
	\end{aligned}
	\end{equation}
	which approximates the true posterior on ${{\textbf{Z}}_{e}}$. In Eq. \ref{eq6}, ${{\theta }_{e}}$ denotes the weights of the encoder in auto-encoder neural network architecture.
	Latent representations ${{\textbf{Z}}_{e}}=[{{z}_{{{e}_{1}}}},{{z}_{{{e}_{2}}}},....,{{z}_{{{e}_{N}}}}]\in {{\mathbb{R}}^{N\times L}}$ are also sampled employing the re-parameterization method \cite{kingma2013auto}, ${{z}_{{{e}_{n}}}}={{\mu }_{e}}({{y}_{n}})+{{\upsilon }_{n}}\odot \sigma _{e}({{y}_{n}})$,
	%
	%
	%\begin{equation}\label{eq9}
	%\begin{aligned}
	%{{z}_{{{e}_{n}}}}={{\mu }_{e}}({{y}_{n}})+{{\upsilon }_{n}}\odot \sigma %_{e}({{y}_{n}}), 
	%\end{aligned}
	%\end{equation}
	%
	%
	%
	where $\odot$ denotes the element-wise product and $\upsilon$ is a random number drawn from a normal distribution.
	We compute the resulting evidence lower bound (ELBO) as:
	\begin{equation}\label{eq7}
	\begin{aligned}
	& \log p(\textbf{Y}|\textbf{X},\textbf{W},{{\theta }_{d}},{{\theta }_{\mathcal{G}\mathcal{P}}})\ge {{\mathbb{E}}_{Z\sim {{q}_{{{\theta }_{e}}}}}}\left[ \sum\limits_{n}{\log \mathcal{N}({{y}_{n}}|{{f}_{d}}({{z}_{n}}))+\log p({{\textbf{Z}}_{e}}|\textbf{X},\textbf{W},{{\theta }_{\mathcal{G}\mathcal{P}}})} \right] \\ 
	& +\frac{1}{2}\sum\limits_{nl}{\log (\sigma _{q}^{2}{{({{y}_{n}})}_{l}})}+const.
	\end{aligned}
	\end{equation}
	To increase  the ELBO as much as possible, we apply stochastic backpropagation \cite{kingma2013auto}. Individually, we approximate the expectation by sampling from a reparameterized variational posterior over the latent representations, achieving the resulting loss function:
	\begin{equation}\label{eq8}
	\begin{aligned}
	& \mathcal{L}({{\theta }_{d}},{{\theta }_{e}},{{\theta }_{\mathcal{G}\mathcal{P}}})= \\ 
	& \sum\limits_{n}{\underbrace{\frac{{{\left\| {{y}_{n}}-{{f}_{d}}({{z}_{{{e}_{n}}}}) \right\|}^{2}}}{2\sigma _{y}^{2}}}_{L2\ reconstruction\ loss}}-\underbrace{\log p({{\textbf{Z}}_{e}}|\textbf{X},\textbf{W},{{\theta }_{\mathcal{G}\mathcal{P}}})}_{\mathcal{G}\mathcal{P}}+\underbrace{\frac{1}{2}\sum\limits_{nl}{\log (\sigma _{{{z}_{e}}}^{2}{{({{y}_{n}})}_{l}})}}_{regularization} \\ 
	& +NK\log \sigma _{y}^{2} \\ 
	\end{aligned}
	\end{equation}
	where we optimize regarding ${{\theta }_{d}}$, ${{\theta }_{e}}$, and ${{\theta }_{\mathcal{G}\mathcal{P}}}$. 
	We optimize loss function through Adam optimizer with a learning rate of 0.001. We experimentally noted that minimizing loss function was developed by first training the encoder and the decoder within the VAE, next optimizing the GP weights by frozen encoder and decoder for 100 epochs (the learning rate of 0.01), last, optimizing all parameters jointly in our MGP-VAE model.  
	\subsection{Missing Modality Imputation}
	We derive an approximate predictive posterior for MGP-VAE that enables missing modality predictions of high-dimensional samples. Specifically, given training samples $Y$, subject feature vectors $X$, and modality feature vectors $W$ , the prediction for the missing data ${y}_{t}$ of subject ${p}_{t}$ in modality  ${m}_{t}$ is given by:
	\begin{equation}\label{eq10}
	\begin{aligned}
	& p({{y}_{t}}|{{x}_{t}},{{w}_{t}},\textbf{Y},\textbf{X},\textbf{W})= \\ 
	& \int{\underbrace{p({{y}_{t}}|{{z}_{t}})}_{decoding\ missing\ data}}\underbrace{p({{z}_{t}}|{{x}_{t}},{{w}_{t}},{{\textbf{Z}}_{e}},\textbf{X},\textbf{W})}_{\mathcal{GP}\ prediction\ of z_t}\underbrace{q({{\textbf{\textbf{Z}}}_{e}}|\textbf{Y})}_{encoding\ all\ training\ data}d{{z}_{t}}d\textbf{Z}_{e}
	\end{aligned}
	\end{equation}
	where ${{x}_{t}}$ and ${{w}_{t}}$ are feature vectors of subject ${{p}_{t}}$ and sub-modality ${{m}_{t}}$, respectively. The approximation in Eq. \ref{eq10} is achieved by substituting  the exact posterior on ${{\textbf{Z}}_{e}}$ with the variational distribution $q({{\textbf{Z}}_{e}}|\textbf{Y})$ (see \cite{casale2018gaussian}). According to Eq. \ref{eq10}, the missing sub-modality can be computed by the three steps. First, we encode all training image data in the latent space by the encoder, ${\textbf{Z}}_{e}$. Next,  predict latent representation ${{z}_{t}}$ of image ${{y}_{t}}$ through the GP model using $m$, $\textbf{X}$, and $\textbf{W}$. Lastly, latent representation ${z}_{t}$ is decoded to the high-dimensional image space through the decoder as missing modality imputation.

\textbf{3D Variational Encoder-Decoder Network Architecture.}
The encoder part employs four spatial levels, where each level consists of two convolution layers with $3\times3\times3$ kernel and ELU. The first convolution layer is without downsampling (stride = 1), while the second one applies strode convolution for downsizing. We follow a typical VAE approach to downsize image dimensions by two progressively, but with fixed feature size equal to 32 except encoder endpoint with 16 feature maps. The encoder endpoint has size $16\times4\times4\times4$, followed by a fully connected layer, and is 16 times spatially smaller than the input volume of $64\times64\times64$. The decoder structure is similar to the encoder one, but each level begins with volumetric upsampling using the nearest neighbor algorithm.
	\section{Experiments and Results}
	\subsubsection{Data.}
%	\subsection{Data}
	We assess our method on the training set of BRATS'19 \cite{Menze2015Brain,bakas2017advancing,bakas2018identifying}, which includes the scans of 335 patients.
	Each subject is scanned with four T1, T1c, T2, and F sequences. All scans are skull-striped and re-sampled to an isotropic 1mm resolution, and four sequences of each patient have been co-registered. Radiologists provided the ground truth labels. The segmentation comprise the following tumor tissue labels: 1) non-enhancing tumor, 2) edema, 3) enhancing core.
	Implementation of the MGP-VAE is available\footnote[1]{\url{https://github.com/hamghalam/MGP-VAE}}.
	\begin{figure}[!t]
		\centering
		\includegraphics[width=10.5cm]{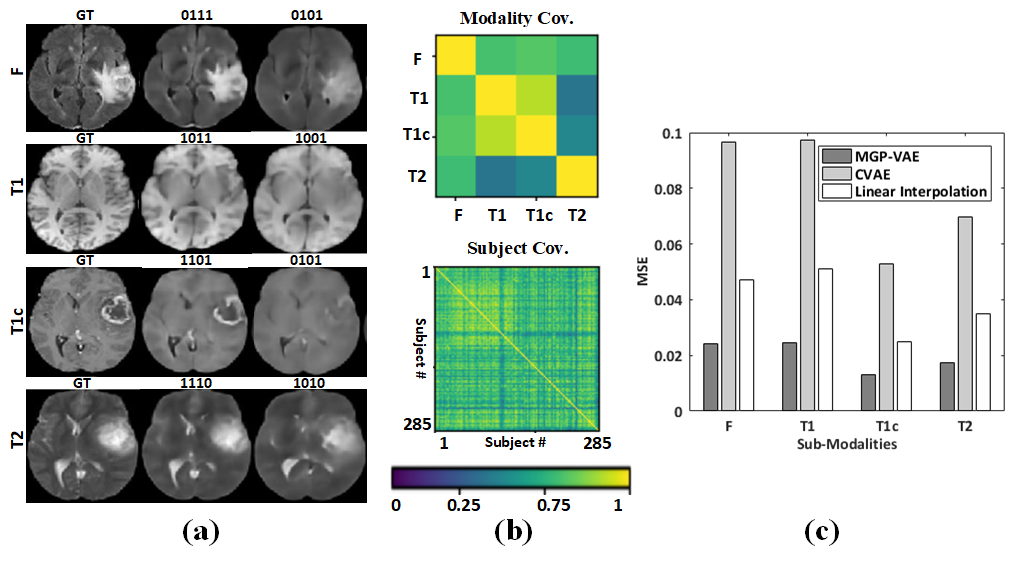} %fig2.png 
		\vspace{-.45cm}
		\caption{(a) Example of modality completion (each row corresponds to a particular sub-modality) given a subset of sub-modalities as input. The 4-bit strings on top of each slice determine present and absent sub-modality with 1 and 0, respectively (bit order from left to right F, T1, T1c, and T2). (b) Covariances between sub-modalities and subjects are modeled through a GP prior model. (c) We compared our method with CVAE and VAE, the baseline of HVED and other well-known imputation methods.}		
		\label{fig:example}
		\vspace{-.4cm}		
	\end{figure}
%
%\vspace{-.4cm}
	%
	%
%
%	
%\textbf{Missing Modality Imputation.}
\subsection{Missing Modality Imputation}
	Fig. \ref{fig:example}(a) illustrates a qualitative evaluation of each sub-modality reconstruction with one missing sub-modality (second column) and two missing sub-modalities (third column). Our model proposes to reconstruct the brain and tumor tissue even when the tumor information is missing or not clear by coupling information from available samples. Comparing reconstructions using two sub-modalities and three sub-modalities confirms that the reconstructed volumes preserve high-frequency details. This suggests that the MGP-VAE model can effectively learn relations between available sub-modalities in different subjects  (Fig. \ref{fig:example}(b)). 	
The PSNR values for imputation are F = 27.95, T1 = 27.80, T1c = 29.43, and T2 = 27.99 based on three available sub-modalities. Similarly, we have  F = 22.36, T1 = 22.56, T1c = 24.86, and T2 = 22.66 with two available sub-modalities. Besides, Mean Squared Error (MSE) for each sub-modality is considered as an evaluation metric to compare MGP-VAE with CVAE and linear interpolation. The latter applies linear interpolation between available sub-modalities of a subject in the latent space learned through VAE to predict the missing sequence (Fig. \ref{fig:example}(c)). The CVAE indeed improves VAE in image generation by conditioning the encoder and decoder to the desired input. However, when we have missing input data, CVAE has a confined ability to create the latent variable for unseen data compared to VAE. This might be because CVAE is more restricted to learn particular data features (latent representation) from observed input data. Therefore it has dedicated latent variables with limited features from missing data. The latent space of VAE contains more general characteristics which can be used to predict missing data.
\subsection{Glioma Segmentation}
\begin{figure}
		\centering
		\includegraphics[width=5.8cm]{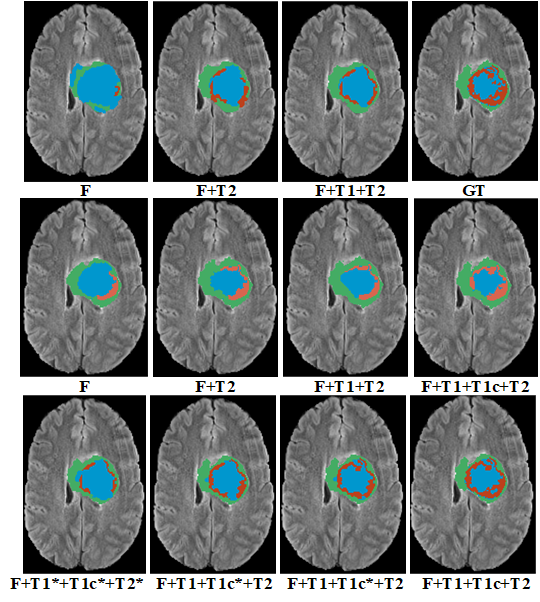} %width=12cm
		\vspace{-.2 cm}
		\caption{The first row explains the effects of different combinations of input sub-modalities overlaid on the FLAIR(F) slice without the imputation. The second row illustrates the HVED results. The last row represents the segmentation with imputation based on MGP-VAE method. The segmentation colors describe edema (green), non-enhancing (blue), and enhancing (red). * indicates imputed sub-modalities.}		
		\label{fig:seg}	
		\vspace{-.4cm}	
	\end{figure}

%\textbf{Glioma Segmentation.}
To assess the MGP-VAE, we examine it on the brain tumor segmentation framework and compare it with two state-of-the-art methods for all the possible subset of sub-modalities. The first, HVED \cite{dorent2019hetero} is the state-of-the-art method based on VAE for brain tumor segmentation with missing sub-modalities. The second approach, HeMIS \cite{havaei2016hemis}, combines the available sub-modalities based on feature maps moments. We adopt the 3D U-Net architecture \cite{cicek20163d,ronneberger2015u} to segment glioma where the available sub-modalities and imputed ones are concatenated as input multi-modal MRI scans. We use the Dice score to measure segmentation accuracy in clinically significant glioma subregions: WT, TC, and ET in Table \ref{tab:missing_cmp_others}. We have almost the same performance in Table \ref{tab:missing_cmp_others} if all sub-modalities are available (without imputation). Our method is designed and optimized to address problems where either one, two, or three of four sub-modalities may be missing. When all the sub-modalities are available, this is a different scenario \cite{hamghalam2020high,hamghalam2020highj,soleymanifard2019segmentation,hamghalam2020transforming,hamghalam2019convolutional,hatami2019}. Moreover, Fig. \ref{fig:seg} shows comparative segmentation results of the BraTS'19 dataset without (first row) and with (second row) imputation through MGP-VAE. The last column (first row) is the ground truth.
	\begin{table*}%[!t]
		\centering
		\caption{Comparison of MGP-VAE model with HeMIS and HVED model (Dice $\%$) for all subset of available sub-modalities. Sub-modalities present are denoted by 1, the missing ones by 0. The IQR is the interquartile range (IQR) and * indicates significant improvement by a Wilcoxon test ($p<0.05$).} 
		\vspace{-.3 cm}
		\label{tab:missing_cmp_others}
		\begin{tabular}{@{\extracolsep{\fill}}cccc|ccc|ccc|ccc}
			\toprule[1pt] 
			%\rule[-1ex]{0pt}{2.5ex}
			\multicolumn{4}{c|}{\textbf{Modalities}} &\multicolumn{3}{c|}{\textbf{WT}} & \multicolumn{3}{c|}{\textbf{TC}} & \multicolumn{3}{c}{\textbf{ET}} \\
			%\cmidrule[\heavyrulewidth]{1-5} 
			%\cmidrule[\heavyrulewidth]{7-9}
			\midrule
			F&T1&T1c&T2&\textbf{HeMIS} &\textbf{HVED} & \textbf{Ours} & \textbf{HeMIS} &\textbf{HVED} & \textbf{Ours} &\textbf{HeMIS} &\textbf{HVED} & \textbf{Ours}  \\
			\midrule
			0 &0 &0 &1 &78.0 &79.9 &\textbf{81.1*} &49.3 &52.8 &\textbf{56.2*} &22.1 &29.9 &\textbf{30.8}\\
			0 &0 &1 &0 &57.6 &61.7 &\textbf{63.2*} &57.7 &65.8 &\textbf{68.1*}&59.5 &65.1 &\textbf{66.4*}\\
			0 &1 &0 &0 &\textbf{53.3} &51.5 &53.2 &37.0 &36.2 &\textbf{39.9*} &11.3 &13.2 &\textbf{14.2}\\
			1 &0 &0 &0 &78.9 &81.0 &\textbf{83.3*} &48.8 &49.9 &\textbf{52.7*} &24.2 &24.1 &\textbf{25.2}\\
			0 &0 &1 &1 &80.1 &81.8 &\textbf{83.1*} &68.3 &73.1 &\textbf{75.7*} &67.5 &69.2 &\textbf{70.5*}\\
			0 &1 &1 &0 &62.6 &66.0 &\textbf{67.6*} &63.4 &69.1 &\textbf{71.8*} &64.3 &66.6 &\textbf{68.0*}\\
			1 &1 &0 &0 &82.9 &83.2 &\textbf{84.7*} &\textbf{56.0} &54.4 &55.3 &\textbf{28.2} &24.1 &25.1\\
			0 &1 &0 &1 &79.8 &81.1 &\textbf{82.0} &52.5 &56.6 &\textbf{58.4*} &27.1 &30.3 &\textbf{31.8*}\\
			1 &0 &0 &1 &84.8 &86.5 &\textbf{87.9*} &58.0 &58.9 &\textbf{61.0} &26.9 &33.7 &\textbf{35.0*}\\
			1 &0 &1 &0 &82.2 &84.8 &\textbf{85.8} &66.7 &72.6 &\textbf{74.6*} &67.0 &\textbf{70.4} &70.2\\
			1 &1 &1 &0 &83.9 &85.6 &\textbf{87.4*} &69.9 &73.6 &\textbf{75.8*} &68.8 &70.2 &\textbf{71.5*}\\
			1 &1 &0 &1 &85.9 &87.1 &\textbf{88.3} &60.1 &61.0 &\textbf{63.0*} &32.3 &33.3 &\textbf{34.5*}\\
			1 &0 &1 &1 &85.9 &87.5 &\textbf{89.0*} &71.6 &75.2 &\textbf{77.4*} &68.9 &70.3 &\textbf{71.4*}\\
			0 &1 &1 &1 &81.3 &82.3 &\textbf{83.6} &69.9 &74.5 &\textbf{76.4*} &68.7 &70.4 &\textbf{71.6*}\\
			\midrule
			\multicolumn{4}{c|}{\textbf{Median}} &80.7 &82.05 &83.45 &59.05 &63.4 &65.55 &45.9 &49.4 &50.7\\
			\multicolumn{4}{c|}{\textbf{IQR}} &9.98 &9.40 &9.80 &17.0 &19.23 &19.75 &41.58 &41.78 &41.33\\
			
				\midrule
				\vspace{-.4 cm}	
		\end{tabular}
	\end{table*}
	\section{Conclusion}
	We have introduced MGP-VAE to predict missing data of subjects in specified MRI sub-modalities using a specialized VAE. Our model incorporates a GP prior over the encoded representation of available volumes to compute correlations among available sub-modalities and subjects. We also validated the robustness of the method with all possible missing sub-modality scenarios on glioma tumors and achieved state-of-the-art segmentation results. Finally, our method offers promising insight for leveraging large but incomplete data sets through one single model. Possible future work focuses on extending the method to various imaging modalities (MRI, PET, and CT) as well as genetic data.
	\section{Acknowledgements}
	This work was funded in part by National Institutes of Health R01 CA233888.
	
	\clearpage
	\bibliographystyle{splncs04}
	\bibliography{mybib_miccai2020_abs}

\end{document}